\documentclass[12pt,preprint]{aastex}
\usepackage{emulateapj5}

\def\fmag{\hbox{$.\!\!^m$}}
\newcommand{\mincir}{\raise
-2.truept\hbox{\rlap{\hbox{$\sim$}}\raise5.truept\hbox{$<$}\ }}
\newcommand{\magcir}{\raise
-2.truept\hbox{\rlap{\hbox{$\sim$}}\raise5.truept\hbox{$>$}\ }}

\newenvironment{inlinefigure}{%
\def\@captype{inlinesfigure}%
\noindent\begin{minipage}{\linewidth}\begin{center}}
{\end{center}\end{minipage}\smallskip}
\makeatother

\begin{document}

\title{The Relation between Morphology and Dynamics of Poor Groups of Galaxies}

\author{Hrant M. Tovmassian\altaffilmark{1}, and M.Plionis\altaffilmark{1,2}}

\altaffiltext{1}{Instituto Nacional de Astrof\'{\i}sica \'Optica y 
Electr\'onica, 72840, Puebla, Pue, Mexico, email: hrant@inaoep.mx}

\altaffiltext{2}{Institute of Astronomy \& Astrophysics, National 
Observatory of Athens, I.Metaxa \& B.Pavlou, P.Penteli 152 36, 
Athens, Greece, e-mail:mplionis@astro.noa.gr} 

\begin{abstract}

We investigate the relation between the projected morphology and the
velocity dispersion of groups of galaxies using two recently compiled
group catalogs, one based on the 2MASS redshift
survey (Crook et al. 2007) and the other on the SDSS Data Release 5
galaxy catalog (Tago et al. 2008).
We analyse a suitable subsample of groups from each catalog selected such that
it minimizes possible systematic effects.
We find that the velocity dispersion of groups is strongly correlated with the group 
projected shape and size, with elongated and larger groups having a lower velocity 
dispersion. Such a correlation could be attributed to the dynamical evolution of groups, 
with groups in the initial stages of formation, before virialization is complete, 
having small velocity dispersion, 
a large size and an elongated shape that reflects the anisotropic accretion of galaxies 
along filamentary structures. However, we show that the same sort of correlations could also be
reproduced in prolate-like groups, irrespective of their dynamical state, if 
the net galaxy motion is preferentially along the group elongation, since then the groups 
oriented close to the line of sight will appear more spherical, will
have a small 
projected size and high velocity dispersion, while groups oriented close to the 
sky-plane will appear larger in projection, more elongated, and will have smaller 
velocity dispersion. Although both factors must play a role in shaping the
observed correlations, we attempt to disentangle them by performing
tests that relate only to the dynamical evolution of groups (ie.,
calculating the fraction of early type
galaxies in groups and the projected group compactness).
Indeed we find a strong positive (negative) correlation between the 
group velocity dispersion (group projected major axis) with the fraction
of early type galaxy members. 
We conclude
that (a) the observed dependencies of the group velocity dispersion on 
the group projected size and shape, should be attributed
mostly to the dynamical state of groups and (b) groups of galaxies in the local universe
do not constitute a family of objects in dynamical equilibrium, but rather a family of cosmic
structures that are presently at various stages of their virialization
process.
\end{abstract}


\section{Introduction}

Poor groups of galaxies are the first level structures of galaxies in the 
hierarchy of cosmic structure formation. Considerable effort has been 
put in identifying such objects in redshift surveys of galaxies and it 
has been found that a large number of galaxies in the local Universe 
are indeed members of such groups (e.g. Huchra \& Geller 1982; Tully 1987; 
Nolthenius \& White 1987; Ramella et al. 2002; Merch\'an \& Zandivarez 
2002, 2005; Gal et al. 2003; Gerke et al. 2004; Lee et al. 2004; 
Lopes et al. 2004; Eke et al. 2004; Tago et al. 2006, 2008; 
Berlind et al. 2006; Crook et al. 2007; Yang et al. 2008). 

The determination of the dynamical state 
and evolution of groups is an important step for investigating 
the hierarchical galaxy formation theories. 
Many dynamical and morphological studies have been restricted to 
compact groups (e.g. Kelm \& Focardi 2004, Da Rocha, Ziegler, \& 
Mendes de Oliveira 2007; Coziol \& Plauchi-Frayn 2007). The intrinsic elongated
(mostly prolate-like) 
shape of groups (e.g, Hickson et al. 1984; Malykh \& Orlov 
1986; Orlov, Petrova, Tarantaev 2001; 
Plionis, Basilakos \& Tovmassian 2004; Plionis, Basilakos, \& 
Ragone-Figueroa 2006; Wang et al. 2008) is a very important factor for the determination
 of their dynamical state (see however Robotham, Phillips \& de
 Propris 2007). Tovmassian, Martinez \& Tiersch (1999) 
showed that the projected length and velocity dispersion of Hickson 
compact groups (Hickson 1982) are anti-correlated, and suggested that
 member galaxies in these groups possibly move preferentially along 
the group major axis in quasi-stable orbits (see also Tovmassian 2001, 2002).
Using the Millenium simulation, Diaz-Gimenez et al. (2008) have found
a weak correlations between projected elongation and line of
sight velocity dispersion in physically dense compact groups, but
also in groups characterized as compact due to chance alignments along the line-of-sight.
%, point to sample-selection as a paussible cause of these correlations.

Tovmassian \& Chavushyan (2000) and Tovmassian, Plionis \& 
Torres-Papaqui (2006) have found that members of loose groups in which 
compact groups are embedded, appear to move in similar elongated orbits 
around the common gravitational center of the corresponding group. However, 
such groups may represent only a special class and the study of generic 
poor groups is important for our understanding of their formation and 
evolution processes. 

In this paper we study the relation between group morphology and dynamics, using
as a measure of group morphology the projected axial ratio of the fitted ellipse and the
projected group size, while as a measure of the group dynamical state we 
use its velocity dispersion and galaxy morphological content. 
%For the purpose of this study we use the 
%groups of the 2MASS High Density Contrast (HDC) catalog (Crook et al. 2007), 
%and the group catalog based on the SDSS Data Release 5 (Tago et al. 2008).

\section{Data and Methods}
\subsection{Group Sample Selection}
In order to investigate group dynamics it is important to use groups not contaminated, as much
as possible, by field galaxies. We want to stress that random 
projection of field galaxies over groups could affect their true shape and dynamical
 parameters, and also their morphological content. It is obvious that the probability 
of a group being significantly affected by random projections is
inversely proportional to the group 
galaxy membership, $n_m$. Projection of even one field galaxy may significantly alter 
the dynamical parameters of poor groups consisting of a few members. 
For example, Ramella et al. (2002) mention that 20\%-60\% of their groups 
which mainly consist of less than ten galaxies, are expected to be contaminated by 
superpositions of field galaxies. 
Furthermore, the effect of discreteness in the determination of the shape of groups 
with a few members is severe (eg. Paz et al. 2006; Plionis et al. 2006), a fact 
which results into artificially elongated shapes.
Moreover, Robotham et al. (2007) 
claimed that poor groups with few galaxy members may have an oblate configuration. The 
expected galaxy orbits and therefore dynamics in oblate groups differ from that of 
groups with a prolate-like configuration, which has been shown to be the dominant
group and cluster shape (eg., Malykh \& Orlov 1986; Plionis, Barrow \&
Frenk 1991; de Theije, Katgert \& van Kampen 1995;
Basilakos, Plionis \& Maddox 2000; Cooray 2000; Plionis et
al. 2004; 2006; Sereno et al. 2006; Wang et al. 2008).

In this study we use the 2MASS High Density Contrast (HDC) group 
catalog (Crook et al. 2007) which was constructed by a friends-of-friends algorithm
(eg. Huchra \& Geller 1982) such that the groups 
correspond to an overdensity $\delta\rho/\rho \ge 80$. We have choosen
to study this catalogue and not the lower density contrast (LDC) one, 
which is based on $\delta\rho/\rho\ge 12$, since
we believe that the HDC catalog
is less prone to projection, interloper contamination and
contamination by the large-scale structures from which galaxies are
accreted to the groups.
We also use the group catalog constructed by Tago et al. (2008) from the 
SDSS Data Release 5, which from now on
we will tab as ``Tago-SDSS''. Although the authors do not provide the
overdensity threshold to which their groups correspond, a crude
calculation gives $\delta\rho/\rho\simeq 260$.
The overdensity difference between the two group catalogues
is reflected in their projected size distribution with the ``Tago-SDSS'' groups 
being significantly smaller than the 2MASS-HDC groups,
as can be verified inspecting figures 3-6 and 8 further below.

Taking into account the problems discussed in the beginning of this section
we wish to limit our study to groups with more than eight members ($n_m\geq9$),
and since our aim is to study poor groups we also put an upper limit of $n_m\le 12$. 
Furthermore by studying groups in a small $n_m$ range we significantly reduce
 the variable, due to the different group
 membership, discreteness effects on their measured shape (eg., Paz et al. 2006; Plionis et al. 2006). 

One more important issue regarding mostly
the high $n_m$ groups is the fact that the increasing size of the friend-of-friend 
linking radii (radial and tranverse), 
necessary to take into account the decrease of the selection function 
with redshift, tends to join at higher redshifts nearby (clustered) groups into single entities.
Therefore, the probability that the groups are real dynamical entities, should 
decrease with redshift. 
%So, there is a trade-off between the group multiplicity, 
%limiting distance and good statistic. 
However, both of the group catalogs used (Crook et al 2007; Tago et al. 2008) 
have been constructed taking special care for
this effect and it appears that indeed the artificial trends, found in previous 
group catalogs (see discussion in Plionis et al. 2004; 2006) have been significantly
suppressed. None the less, by visually inspecting the apparently large and 
high velocity dispersion rich groups we have found that in quite a few occasions they 
appear to be the by-product of joining neighboring groups. 
We would like to remind the reader that Rose (1977), Mamon (1986, 2008) and 
Walke \& Mamon (1989) put forward the idea that some ordinary and compact
 groups could well be a result of such projection effect (see relevant
 recent works of Diaz-Gimenez et al. 2008 and Brasseur et al. 2008).

Of the groups with 
$9\le n_m\le 12$ we have found, in the 2MASS-HDC sample, only one that clearly falls in 
this category (No 1218; see figure 1) which we exclude from our analysis. 
Indeed, the projected distribution of members of this group shows that eight of its 
members compose a relatively compact group with mean $V=1713\pm233$ 
km s$^{-1}$, and a wide triplet located at projected distance of 
$\sim 1.4 \; h^{-1}$ Mpc to the south 
with mean $V=1074\pm111$ km s$^{-1}$. Another three of the supposed 
members of this group are located at projected distance of 
$\sim 1 \; h^{-1}$ Mpc to the south of 
the first subgroup (distances are measured at the mean redshift of the
whole group)\footnote{Throughout this work we use $H_0=100 \;h$ km 
s$^{-1}$ Mpc$^{-1}$ with $h=0.73$.}.
As another example,
we present in Figs. 1 the map of 2MASS-HDC group No 384 with a multiplicity $n_m=13$ 
which could consist of two-three probably unrelated groups. 

\begin{inlinefigure}
\centering\leavevmode
\epsscale{1.0} 
\plotone{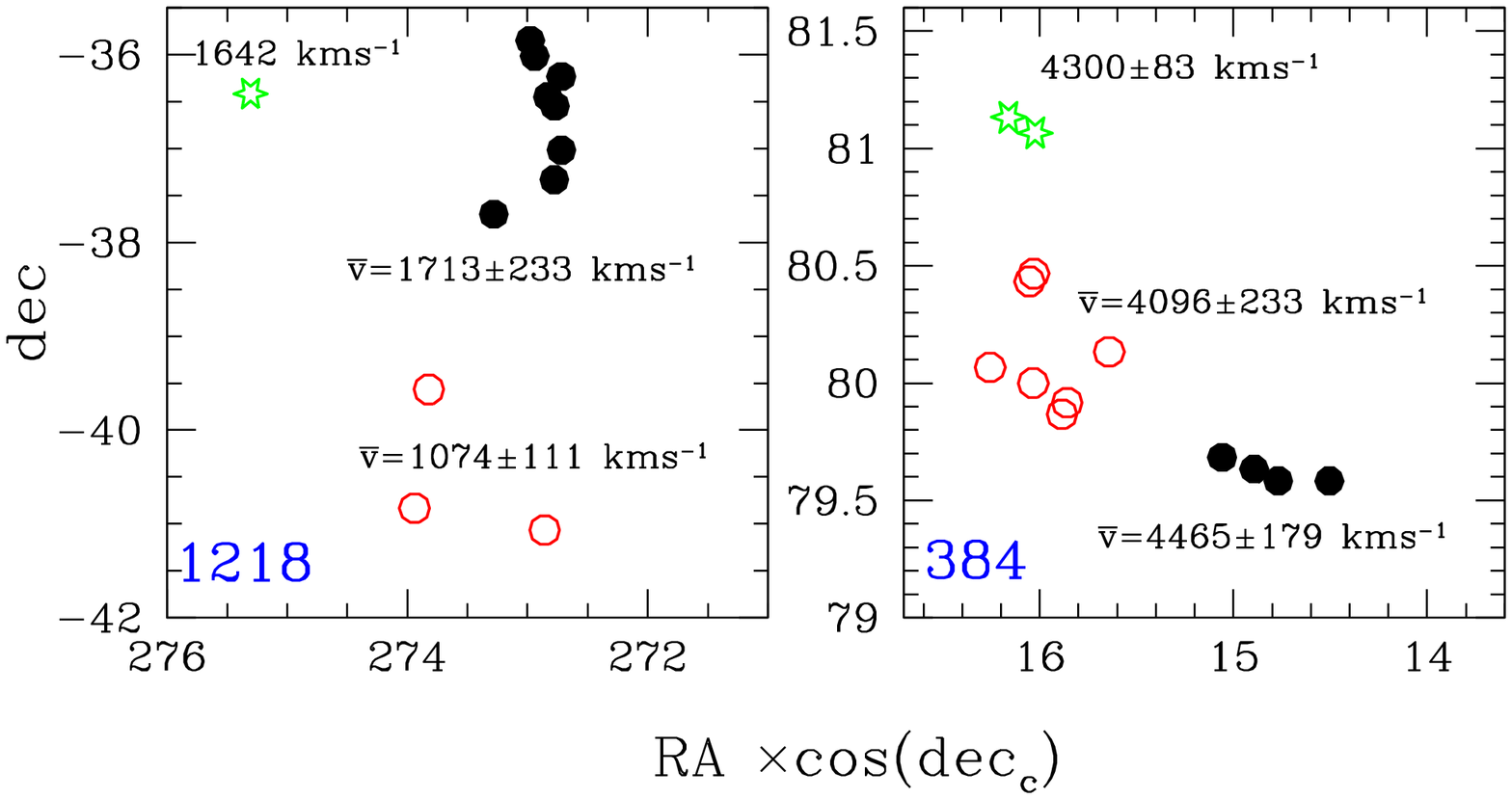}
\figcaption{Equal area maps of two examples of 2MASS-HDC groups
  suspect of being composed of unrelated groups and field galaxies:
{\em Left Panel}: The group 1218, with $n_m=12$, consists 
of two clearly distant sub-groups at a projected separation of $\sim
1.4\; h^{-1}$ Mpc (at the mean redshift of the whole
group) and a mean velocity difference of $\sim 700$ km s$^{-1}$.
This group is the only clearly ``problematic'' group in our $n_m=9-12$ sample. 
{\em Right Panel:} The group 384, with $n_m=13$, possibly consists of 
three separate groups with mean radial velocities 
as indicated in the plot. The projected distance (at the mean redshift of the whole group) 
of the northern and southern subgroups from the central 
one is $\sim 2$ and $\sim 1.3 \; h^{-1}$ Mpc, respectively.}
\end{inlinefigure}

We will also analyse groups (with $n_m\ge 9$) based on their estimated virial masses 
since the magnitude limited
nature of the 2MASS galaxy redshift survey implies that groups of the same multiplicity
but at different redshifts correspond to different intrinsic richness.
Indeed the difference of the $K_{\rm total}$ apparent magnitude (Jarrett et al. 2000) between 
the brightest and faintest galaxy in the 2MASS-HDC groups with, for
example, 9 members is $\approx3\fmag5$ for 
nearby groups with $cz \approx 300$ km s$^{-1}$, and systematically 
decreases to about $1\fmag5$ for distant groups with $cz \approx7000$ 
km s$^{-1}$. This shows that faint group members are missed as a function of increasing 
redshift and thus the distant groups, of apparently the same multiplicity as nearby ones,
are intrinsically richer in members above a given luminosity
threshold. However, if assume that the missed faint galaxies randomly 
sample the distribution of group member velocities, then the mass estimate of the 
groups will not be systematically affected by the omission of these fainter galaxies and 
therefore performing an analysis of low and intermediate mass groups (ie., excluding
the apparently massive systems which could be artificial) circumvents the previously mentioned problems.
Note however that the above assumption may not be valid for virialized
groups in which dynamical friction has played a relatively important
role, and in which case their mass may be underestimated.

In order to have as much as possible a representative sample of the
true underlying local group population and to minimize the above mentioned
redshift dependent systematic biases, we have chosen to study from the
 Tago-SDSS catalog the groups within $z\le 0.043$. 
Note that the 2MASS-HDC group sample is by
construction defined in the local universe ($z \le 0.033$). 

\subsection{Shape and Dynamics measures}
We determine the projected group shape diagonalizing the moments of inertia tensor,
which we construct by weighting each member galaxy by $1/K_{\rm
  total}$, with $K_{\rm total}$ the apparent K-band galaxy magnitude.
 This is done in order to weight more the luminous (and thus 
massive) galaxies, which dominate in shaping the group gravitational potential.

Firstly, the galaxy equatorial positions are transformed into an equal area
coordinate system, centered on the group center of mass (which we
determine using obviously the $1/K_{\rm total}$ weighting scheme).
We then evaluate the moments:
\begin{eqnarray}
I_{11} & = & \sum_{i} w_{i}(r_{i}^{2}-x_{i}^{2}) \nonumber \\
I_{22} & = & \sum_{i} w_{i}(r_{i}^{2}-y_{i}^{2}) \nonumber \\
I_{12} & = & I_{21}=-\sum_{i} w_{i}x_{i}y_{i}
\end{eqnarray}
with $w_{i} (=1/K_{\rm total})$ the statistical weight of each member galaxy 
and $r_i$ the distance of the $i^{\rm th}$ galaxy from the group center of mass. 
Note that because the inertia tensor is symmetric, we have
$I_{12}=I_{21}$. Diagonalizing the inertia tensor
\begin{equation}\label{eq:diag}
{\rm det}(I_{ij}-\lambda^{2}M_{2})=0 \;\;\;\;\; {\rm (M_{2} \;is \; 
2 \times 2 \; unit \; matrix.) }
\end{equation}
we obtain the eigenvalues $\lambda_{1}$, $\lambda_{2}$, from which we
define the principal axial ratio of the configuration under study by:
$q=\lambda_{2}/\lambda_{1} (\equiv b/a)$, with $\lambda_{1}>\lambda_{2}$.
%The corresponding eigenvectors provide the direction of the principal axes.

%The resulting eigenvalues are related to the major ($a$) and minor ($b$) axis of the 
%fitted ellipse, while the axial ratio is defined as the fraction: $q=b/a$.
As a measure of the size of the group we also calculate the mean projected galaxy-galaxy 
separation and a variant (see below) which we use to estimate the group virial mass (only
for the Tago et al. groups, since for the 2MASS-HDC groups the relevant 
values are provided by the catalog). 

The group virial radius, used to determine the group mass, is:
\begin{equation}
R_v = \frac{n_m (n_m-1)}{\sum_{i=1}^{n_m-1} \sum_{j=i+1}^{n_m} 
\left[D_L \tan (\delta\theta_{ij})\right]^{-1}}\;\;,
\end{equation}
where $D_L$ is the luminosity distance of the group (using 
$\Omega_{\Lambda}=0.7$, $\Omega_{\rm m}=0.3$, $h=0.73$), 
$\delta\theta_{ij}$ is the angular $(i,j)$-pair separation.
Using the group velocity dispersion and $R_v$ we can 
estimate the group's virial mass 
%and crossing time 
according to:
\begin{equation}
M_v=\frac{3\pi}{2}\frac{\sigma_v^2 R_v}{G}\;\;\;,\;\;\;\;\; %\tau=\frac{R_v}{\sqrt{3} \sigma_v}\;,
\end{equation}
Note that $R_v$ is significantly smaller than
the maximum or the mean group galaxy-pair separation. In what follows
we will use as the major axis of each group, $a$, its mean galaxy-pair
separation and as its minor axis: $b = a q$.

\section{Group Morphology-Dynamics Relation}
\subsection{Background  framework}
Do we expect to find {\em a priori} any morphology-dynamics relation in a sample 
of self-gravitating groups of galaxies which are in dynamical equilibrium? 
The answer is probably no, since in such a case the velocity dispersion of the group should 
reflect its virial mass, while the group shape should be quasi-spherical since 
relaxation  processes will have (mostly) isotropized the initial anisotropic group phase-space.
If however a group morphology-dynamics relation is found, then it could be due to either 
of two possible causes (or a combination of both): 
\begin{inlinefigure}
\centering\leavevmode
\epsscale{1.02} 
\plotone{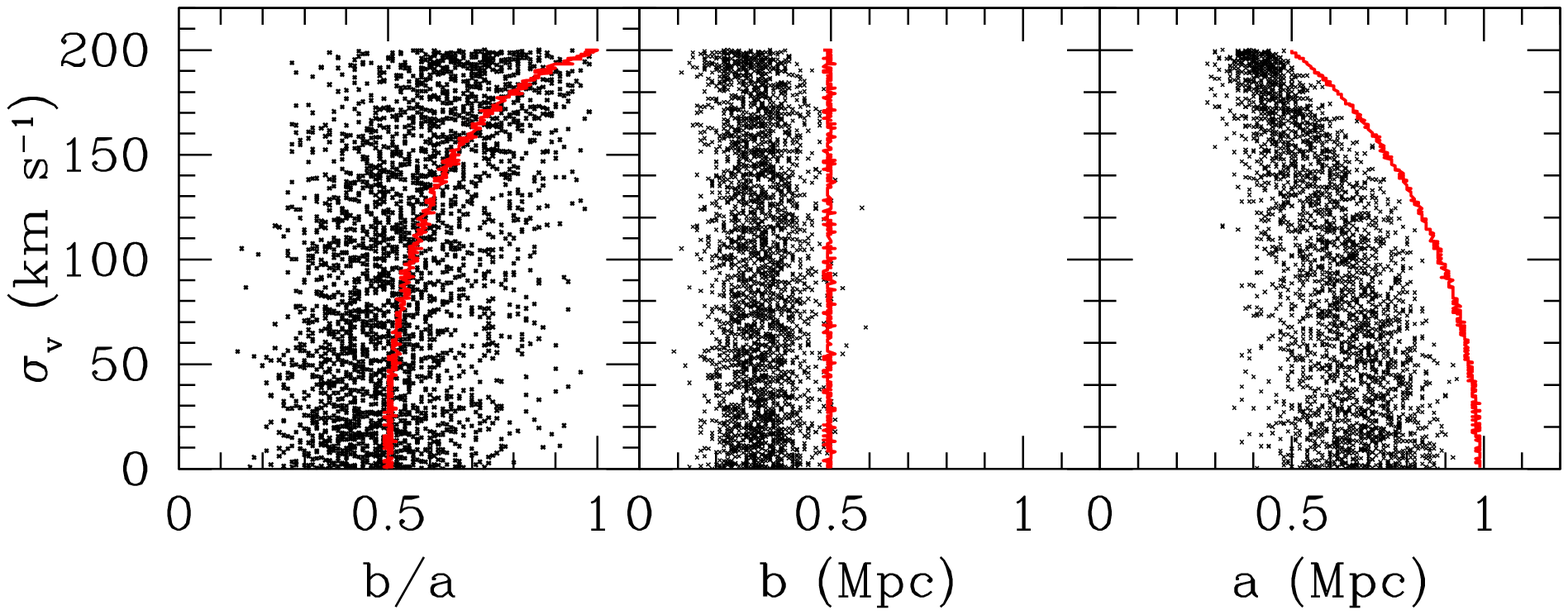}
\figcaption{The expected morphology-dynamics correlations due to the random orientation
with respect to the line-of-sight of prolate spheroids with intrinsic $q=0.5$ and
major axis $a=1$ Mpc. The red line is the expected theoretical curves while the black points
represent random realizations in which each group is sampled by 10 ``galaxies''.}
\end{inlinefigure}

\begin{itemize}
\item The groups of galaxies in the sample are not all virialized but at different
evolutionary stages, and therefore it is possible to find a correlation between their 
velocity dispersion (as a measure of their dynamical state) and the group size and
axial ratio, since a group at its early stages of formation will have a larger 
size, with respect 
to their final dynamical relaxed state, and it will be more elongated reflecting
the initial anisotropic accretion of matter along filamentary structures (eg. West 1994).
\item Since groups of galaxies are prolate-like, as has been found by numerous studies (see
introduction), and if galaxies move predominantly along the group elongation, then 
the size, the axial ratio and the velocity dispersion of the groups will depend on the
group orientation with respect to the line-of-sight.
The nearer is the orientation of the three-dimensional major axis of a group 
to the line-of-sight, the smaller will appear its size, the larger
its axial ratio and velocity dispersion (Tovmassian, Martinez \& Tiersch 1999; 
Tovmassian, Plionis \& Torres-Papaqui 2006, and references therein). In effect, performing
a simple Monte-Carlo simulation in which we randomly orient with respect 
to the line of sight, 1000 intrinsically prolate groups with $q=0.5$,
$a=1$ Mpc and velocity
dispersion $\sigma_v=200$ km s$^{-1}$ strictly along its major axis, we obtain in Fig.2 the red
curves, which constitute the theoretical expectation in the presence of no discreetness.
If now we sample each group by 10 ``galaxies'', we again obtain 
very significant $q-\sigma_v$ and $a-\sigma_v$ correlations but with a large scatter. 
It is important to note that (a) in this model the $a-\sigma_v$ correlation is stronger
than the $q-\sigma_v$ correlation while no $b-\sigma_v$ correlation expected and
(b) the observed scatter in Fig.2 is soley due to sampling randomly 
each Monte-Carlo group by
10 ``galaxies''. More scatter should be expected, however, in a more realistic situation 
in which the intrinsic range of group sizes and velocity dispersions 
would have been taken into account.
\end{itemize}

In fact, the intrinsic spread in group sizes could effectively mask the 
possible morphology-dynamics correlations, 
and thus even a weak correlation should be considered as a
hint of a true underlying effect. 

\subsection{Possible systematic effects}
Any systematic redshift dependence of
$\sigma_v$ and group size, introduced by the convolution of the friends-of-friends
 group finding algorithm and the magnitude-limited nature of the underlying
 galaxy catalogs (see discussion in Plionis et al. 2006), produces a dependence of
the group size and the velocity dispersion with redshift (at higher $z$'s you get large 
groups with higher $\sigma_v$), which could also 
results in an artificial correlation between $a$ and $\sigma_v$, but such that 
$a \propto \sigma_v$ (opposite to what predicted by either of the possibilities
discussed previously).

We have tested whether such bias is present in the presently analysed group samples 
(as has been found in previous group samples; see discussion in Plionis et al. 
2004; 2006) and found 
that for the 2MASS-HDC samples there is a relatively weak $\sigma_v-z$ correlation 
($R=0.29$ with ${\cal P}=0.02$) and 
an insignificant $a-z$ correlation ($R=0.18$ with ${\cal P}=0.18$),
 probably due to the special effort put by the authors 
to reduce such systematics  (Crook et al. 2007). 
The positive $\sigma_v-z$ correlation could well be
due to the expected volume effect (ie., at higher $z$'s a large fraction of the group mass
function is sampled).
Regarding the SDSS-Tago groups we find no correlation whatsoever 
between either $\sigma_v$ nor $a$ with redshift, again an indication that the authors
managed to suppress the systematics from which many other group catalogs suffer.

\subsection{Results}
In Figs. 3-4 we plot the scatter diagrams between the group shape parameters and their 
velocity dispersion for both group catalogs analysed.
It is evident that we do find the qualitatively expected correlations which are
quite strong and significant in both catalogs of groups.
The Pearson correlation coefficients $R$ and corresponding 
random probabilities ${\cal P}$ for the considered group subsamples are presented 
in Table 1. It is evident that the velocity dispersion 
$\sigma_v$ of groups increases with increasing $q$, while it decreases
with increasing group major axis $a$ (while the opposite is expected if it would have
been due to the systematics discussed previously). An additional
systematic effect that acts in the direction of diluting a positive
$q-\sigma_v$ correlation is that, within any $n_m$ bin, the range of
group virial masses traced will tend to
induce an anticorrelation between $q$ and $\sigma_v$, since more massive
halos are more elongated (lower $q$) and have a higher velocity dispersion with
respect to less massive halos (e.g. Jing \& Suto 2002; Kasun \& Evrard 2005;
Allgood et al. 2006; Bett et al. 2007;  Ragone-Figueora \& Plionis
2007; Wang et al. 2008). This implies that the observed shape-dynamics correlation is,
in effect, stronger than what observed.
Note also that a weak but non significant correlation is found between the groups minor axis and 
velocity dispersion. As we will see further below (Fig.4) this is due
to the lower mass groups which do not show a negative $b-\sigma_v$ correlation.

\begin{inlinefigure}
\centering\leavevmode
\epsscale{1.02} 
\plotone{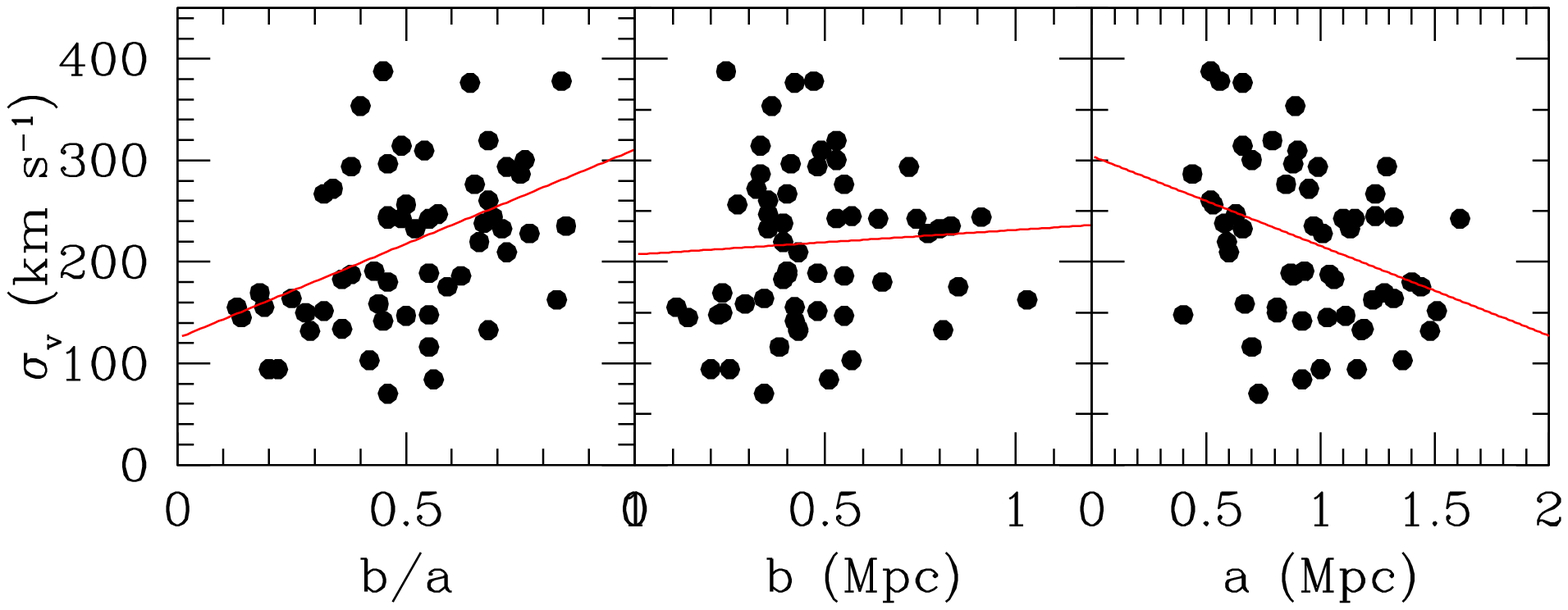}
\figcaption{The morphology-dynamics correlations of the 2MASS-HDC groups: 
From left to right the $q-\sigma_v$, $b-\sigma_v$ and $a-\sigma_v$ scatter diagrams.}
\label{fig4}
\end{inlinefigure}

\begin{inlinefigure}
\centering\leavevmode
\epsscale{1.02} 
\plotone{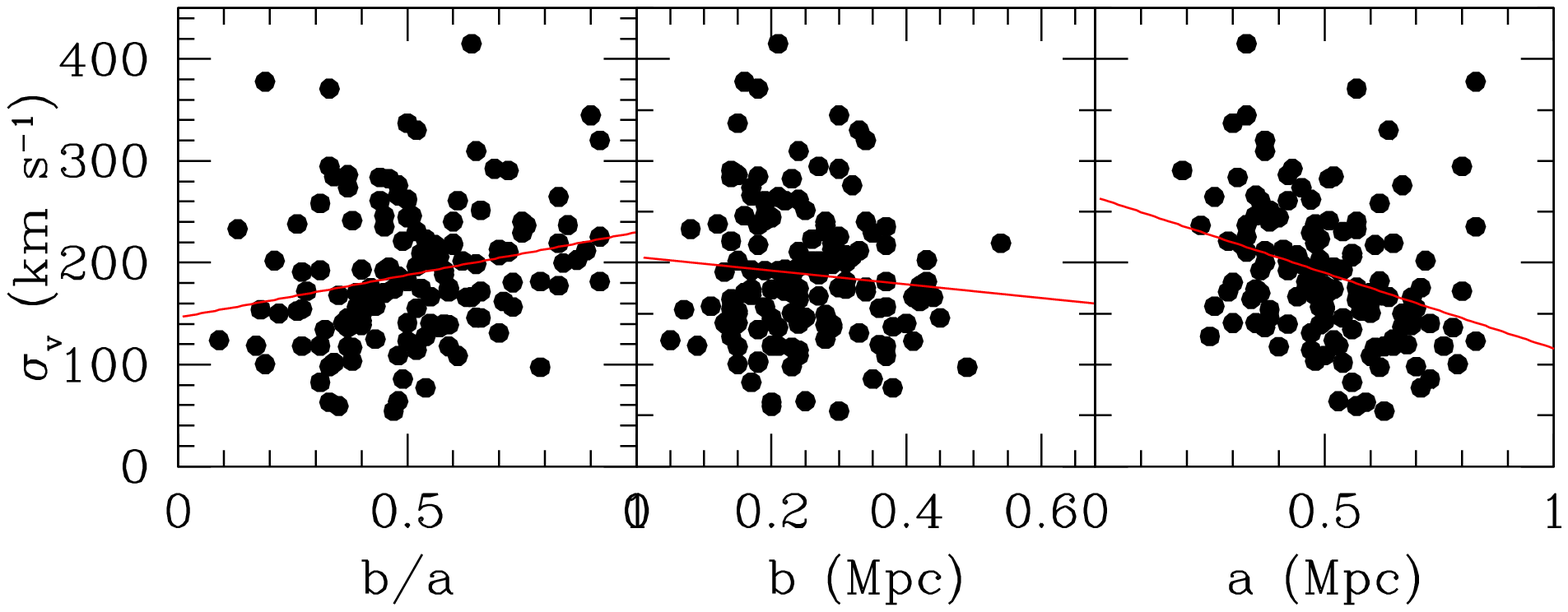}
\figcaption{The morphology-dynamics correlations of the the Tago-SDSS groups: 
From left to right the $q-\sigma_v$, $b-\sigma_v$ and $a-\sigma_v$ scatter diagrams.}
\label{fig6}
\end{inlinefigure}

Probably a more instructive view of the morphology-dynamics correlations, which is free
of the bias of mixing poor nearby and richer distant groups of the same multiplicity ($n_m$),
is to divide the group samples into ranges of mass, according to eq.(4). To this end
we use all groups with $n_m\ge 9$ and divided each group sample in 4 bins of mass, 
excluding groups with $M>10^{14} 
h^{-1} \;M_{\odot}$, which actually correspond to clusters. A further reason to
exclude the high-mass groups is the contamination problem
we have identified in some apparently high velocity dispersion groups
(see Figs. 1).

\begin{inlinefigure}
\centering\leavevmode
\epsscale{1.02} 
\plotone{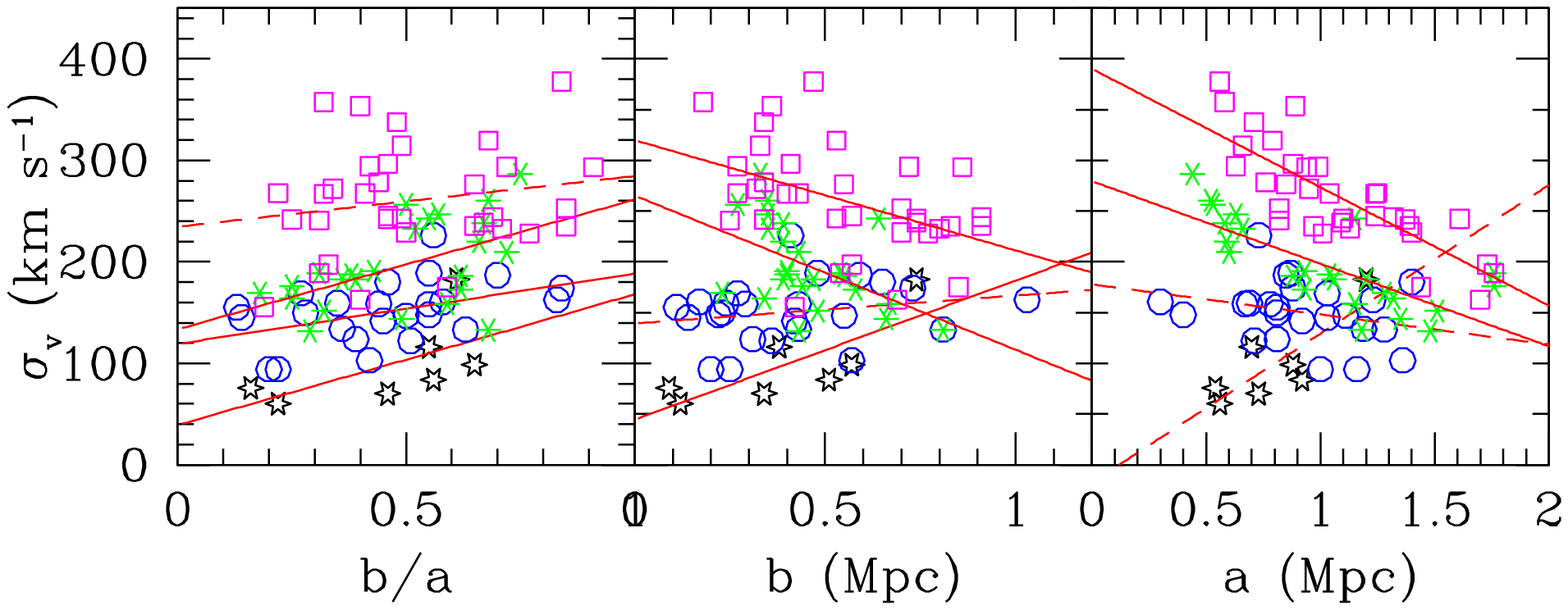}
\figcaption{The morphology-dynamics correlations of the the 2MASS-HDC groups in different
mass ranges: From left to right the $q-\sigma_v$, $b-\sigma_v$ and $a-\sigma_v$ scatter 
diagrams. The stars correspond to $12<\log M/M_{\odot}\le 13$, blue open circles to
$13<\log M/M_{\odot}\le 13.5$, green filled circles to
$13.5<\log M/M_{\odot}\le 13.75$ and magenta open squares $13.75<\log M/M_{\odot}\le 14$.}
\label{fig4}
\end{inlinefigure}

\begin{inlinefigure}
\centering\leavevmode
\epsscale{1.02} 
\plotone{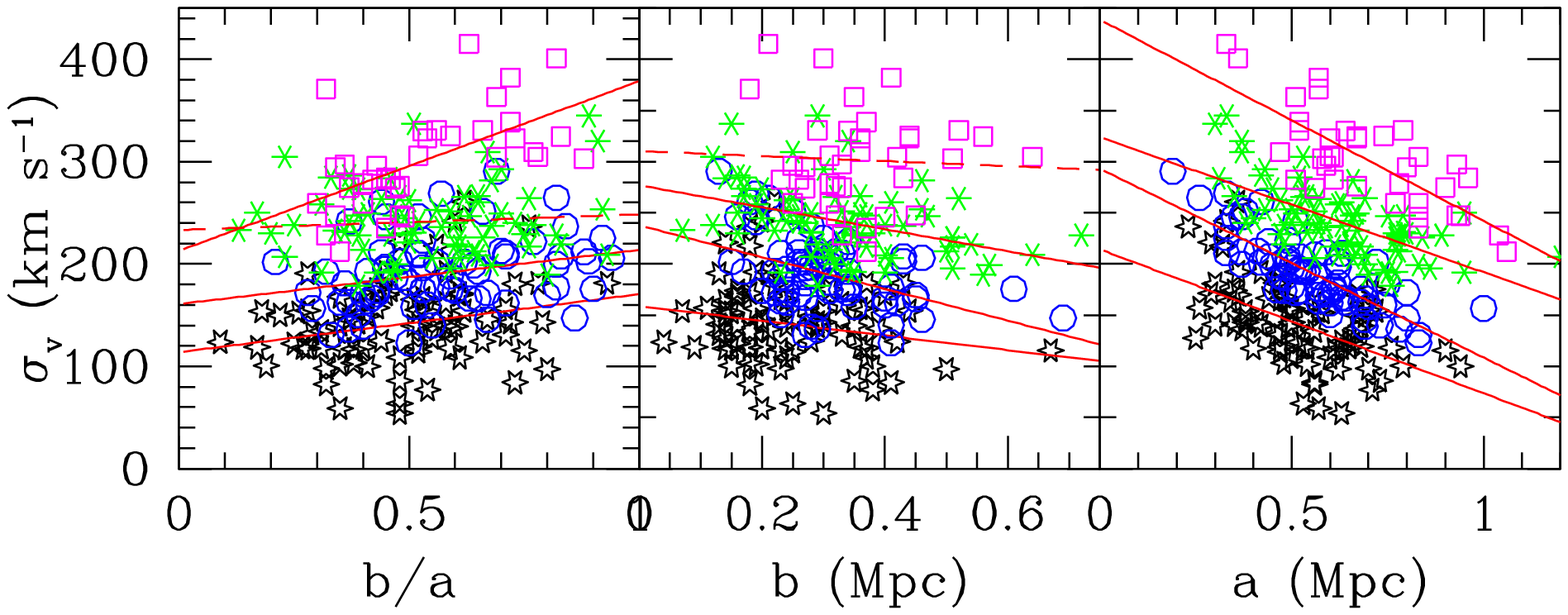}
\figcaption{The morphology-dynamics correlations of the the Tago-SDSS groups in different
mass ranges: From left to right the $q-\sigma_v$, $b-\sigma_v$ and $a-\sigma_v$ scatter 
diagrams. The stars correspond to $12<\log M/M_{\odot}\le 13$, blue open circles to
$13<\log M/M_{\odot}\le 13.25$, green multiple crosses to
$13.25<\log M/M_{\odot}\le 13.5$ and magenta open squares $13.5<\log M/M_{\odot}\le 13.75$.}
\end{inlinefigure}

In Figs 5-6 and in Table 2
we present the same morphology-dynamics correlations as before, but
dividing the groups in 
bins of mass. The correlations are now more evident, with different mass range groups 
occupying a clearly delineated region in the $q-\sigma_v$, $b-\sigma_v$ and $a-\sigma_v$
planes.
Although the number of  2MASS-HDC  groups in each mass range is small, the $q-\sigma_v$
correlations are systematic and significant (with the exception of the highest mass range
which could be affected by the previously mentioned problems). The
$a-\sigma_v$  and $b-\sigma_v$ 
correlations are present only for groups with $M\ge 10^{13.5} \; h^{-1}\; M_{\odot}$.
Similar and more significant correlations are found in the case of the Tago-SDSS groups. 
Here however the $a-\sigma_v$ correlations are extremely significant, while there is
also a significant correlation of the minor axis with velocity dispersion.

\subsection{Orientation or Virialization?}

For the orientation paradigm to work in producing the observed correlations
it is also important to have galaxies moving predominantly along the 
prolate-like group major axis.
Such galaxy motions are generally expected in the hierarchical structure formation 
scenario, where groups and clusters of 
galaxies are formed by anisotropic accretion and merging along filamentary 
large-scale structures (e.g. West 1994). However, the possible predominance 
of such galaxy orbits would also 
imply that the groups are not virialized, since virialization would mix the 
phase-space and erase (mostly) the memory of the initial directional
accretion (see however van Haarlem \& van de Weygaert 1993).

It is interesting to point-out that the observed
$a-\sigma_v$ correlations are generally more significant than those of $q-\sigma_v$, 
a fact which agrees also with the expectations of the orientation
paradigm (see Fig.2), and could be attributed to the dispersion of the minor axis, $b$, which
does not depend on the group orientation.

Therefore, although both discussed causes of the observed group morphology-dynamics
correlations should be at work, we attempt to disentangle which of the
two, if any, dominates. To this end we consider two tests, one based on
the morphological content of groups of galaxies, since a proxy of
their dynamical state should be their morphological content, and the
other based on the expectation of the virialization process to compactify the initial
dispersed group morphology.

\subsubsection{Morphological Content of Groups}
It has been shown that galaxies in 
clusters evolve mainly by dynamical interactions and merging (e.g., 
Goto 2005). It also appears that mergers, strangulation as well as interactions 
with the hot diffuse gas (for the richest groups) act in the group
environment and affect the morphology and gas content of member galaxies
(e.g., Barnes 1985; Zabludoff \& Mulchaey 
1998; Hashimoto \& Oemler 2000; Coziol \& Plauchu-Frayn 2007; Rasmussen et al. 2008). 
The efficiency of galaxy interactions in altering the morphology and
gas content of galaxy group members depends on their relative
velocity, being more efficient in low-velocity dispersion groups (e.g., Mamon
1992), which implies that such processes should be relatively frequent in the
early stages of the group dynamical evolution. 
While galaxy interactions and merging occur,
contemporary the host group evolves dynamically and therefore the fraction 
of E/S0 galaxies should appear high in dynamically advanced (high
velocity-dispersion) groups.
Indeed, the fraction of E/S0 galaxies, $f_{E/S0}$, in groups appears to
increase with increasing group velocity dispersion 
(e.g., Tovmassian, Plionis \& Andernach 2004; Aguerri, Sanchez-Janssen
\& Mu\~noz-Tunon 2007) a fact which could also
be viewed as a manifestation of the known {\em density-morphology} relation at the
groups scale (Postman \& Geller 1984).

%strongly evolve with redshift (e.g., McGee et al. 2008), while it also
%appears to 
\begin{inlinefigure} 
\centering\leavevmode
\epsscale{1.02} 
\plotone{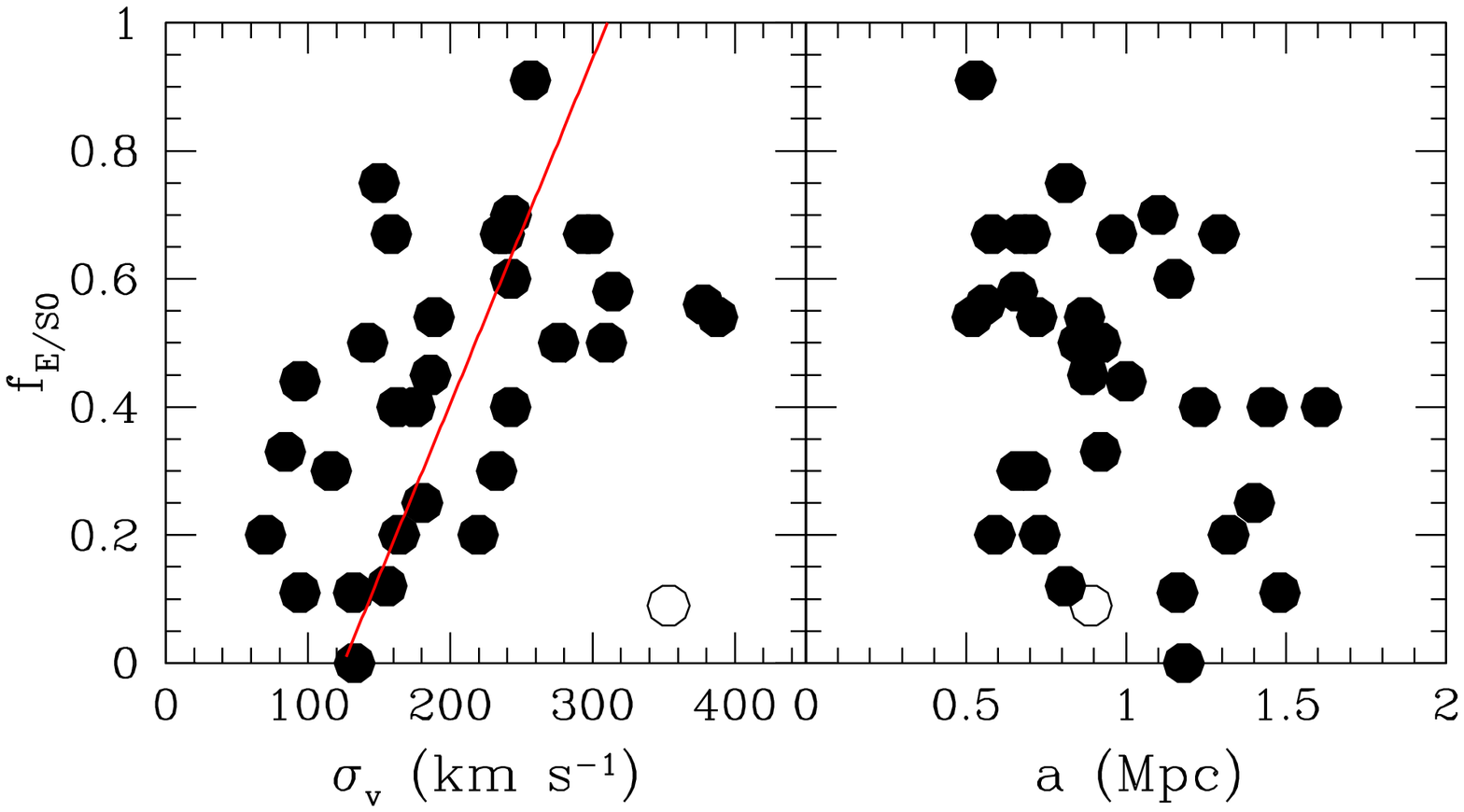} 
\figcaption{Dependence of the fraction $f_{E/S0}$ of early type galaxies 
in groups on the group velocity dispersion $\sigma_v$ (left panel) and major-axis size (right panel).} 
\end{inlinefigure}

Therefore, if we also verify for our current group samples
a correlation between morphological content, $f_{E/S0}$, and group
velocity dispersion then this would clearly suggest that the observed
range of velocity dispersions is related mostly to the group dynamical state
and not to the group orientation (in which case no $f_{E/S0}-\sigma_v$
correlation is expected).
We apply this test to the well defined 2MASS-HDC groups with $n_m=9-12$. We used those 
groups with 9-10 members for which the morphological type of no more than
one member galaxy was unknown, and groups with 11-12 members, the 
morphological types of no more than two galaxies were unknown. We took 
morphological types of member galaxies from the NASA/IPAC Extragalactic 
Database (NED). 
The total number of groups used is 33 and their
 $\sigma_v-f_{E/S0}$ scatter diagram is presented in the left panel of
 Fig. 7 (we also show in this plot with an empty dot the excluded No. 1218 group, suspected
of being contaminated by multipole groups; see section 2.1) and it is
evident that $f_{E/S0}$ strongly increases with increasing $\sigma_v$. 
The Pearson correlation coefficient and random probability are $R=0.54$ and 
${\cal P}=3\times 10^{-4}$, respectively, showing that the $\sigma_v-f_{E/S0}$ 
correlation is indeed very significant. 

During virialization the groups should become more compact and their major axis
should decreases. If this is so, then the  $\sigma_v-f_{E/S0}$ correlation implies
that there should also be a $a-\sigma_v$ correlation. 
Indeed, the right panel of Fig. 7 shows the
corresponding correlation with coefficient $R=0.44$ and random probability ${\cal P}=0.01$.
The $q-f_{E/S0}$ correlation is weak and not significant which could well be due to
the influence of the large dispersion of the group minor axes, $b$. 

\subsubsection{Minor Axes \& Group Projected Size}
In the orientation paradigm, the increase of the velocity dispersion
takes place with corresponding decrease of the projected major axis of 
a group, while the minor axis remain unchanged (see Fig.2). 
Meanwhile, during virialization both the projected major and minor
axis, and thus the projected surface ($S$) 
of a self-gravitating system should decrease becoming more
compact. Indeed, we have found that the projected minor axes of most
of the analysed samples of groups decrease with increasing velocity dispersion. Note however
that projection effects and the presence of interlopers will affect
more the projected minor axis, with respect to the major axis, of an
 intrinsically elongated group,  and thus weaken any true correlations
 between $b$ and $\sigma_v$. This, as well as low-number statistics,
could be the reasons why the low-mass 2MASS-HDC groups (see Fig. 5) 
do not show a negative $b-\sigma_v$ correlation. 

Obviously, the $a-\sigma_v$ and
 $b-\sigma_v$ correlations translate into a $S-\sigma_v$ correlation
with a rate of variation, in the virialization case, which is
higher than that caused by orientation (in which case only $a$
decreases with increasing $\sigma_v$). 
%Figs. 5 and 6 show 
%that the major axis of groups varies by about 3-5 times. 
In Fig. 8 we present the $S-\sigma_v$ correlations 
for the mass-defined 2MASS-HDC and Tago-SDSS groups. It is evident
that, depending on group virial mass, the group surface varies
by 5-8 times within the range covered by the group velocity dispersion. The
 corresponding variation in the case of the orientation paradigm is
expected to be $\sim 1/q$,
which means that for the observed groups (which have $\langle q \rangle \sim 0.5$)
the expected surface variation , within the indicated velocity
dispersion range, is $\sim 2$ (as seen also in right panel of Fig.2), 
significantly smaller than what observed.

\begin{inlinefigure} 
\centering\leavevmode
\epsscale{1.02} 
\plotone{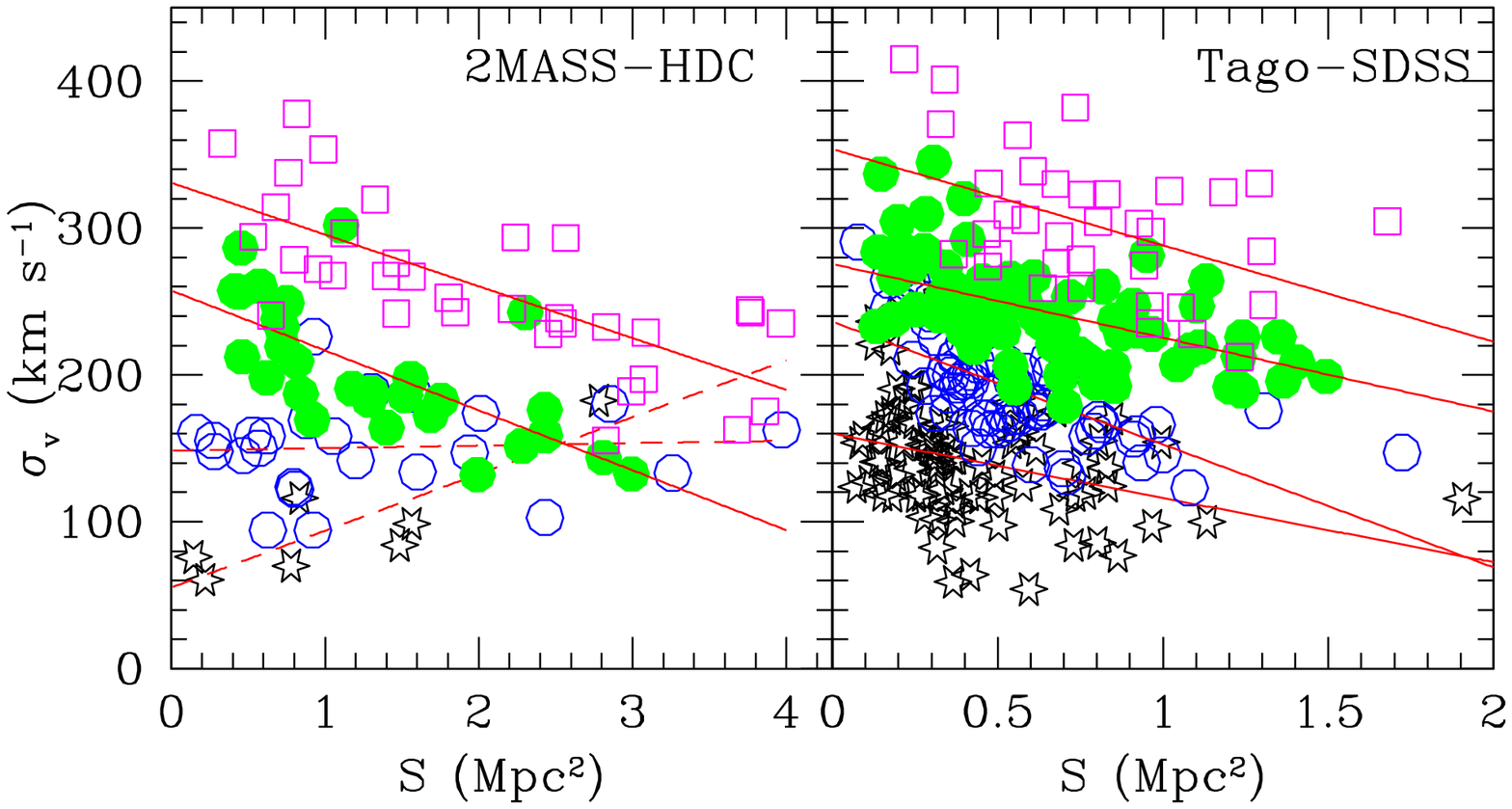} 
\figcaption{Dependence of the projected group surface ($S$)
on the group velocity dispersion, $\sigma_v$ for subsamples of
different mass. {\em Left Panel:} 2MASS-HDC groups (symbols as in Fig. 5) 
and {\em Right Panel:} SDSS-Tago groups (symbols as in Fig.6).}
\end{inlinefigure}

\section{Conclusions}
If a family of cosmic structures are all virialized, there 
is no reason to find any significant morphology-dynamics
correlations.
Such a correlation may be expected in two cases:
(a) if the galaxy group members have a net motion predominantly along the group elongation, 
as expected in dynamical young groups which form by anisotropic accretion 
of matter along filamentary large-scale structures, then due to 
projection there must be a positive correlation between the group 
axial ratio $q$ and the group radial velocity dispersion, and a negative
 correlation between the projected group major axis and the group radial
 velocity dispersion, and (b) if virialization is
 currently at work, which will tend to compactify and sphericalize the initial volume 
from which the structures form, as well as increase its velocity dispersion. 

We searched for such 
correlations using 2MASS-HDC (Crook et al. 2007) and SDSS Data Release
 5 (Tago et al. 2008) catalogs of groups. In order to avoid
 discreteness and interloper contamination effects we have performed two analyses, one
based solely on group samples defined by their apparent multiplicity ($n_m=9-12$) 
and one based on samples defined in bins of group virial mass.

We found significant negative $a-\sigma_v$ and positive $q-\sigma_v$
 correlations, although the observed correlations could have been
substantially weakened by many effects, among which also a 
distance dependent bias by which at larger redshifts
the detected group length and velocity dispersion increases artificially.
However, we have verified that the analysed groups, due to the specific care taken by the authors
that constructed them, do not suffer significantly of this effect.

We have also found a positive and significant correlation between the early type galaxy 
content and the group velocity dispersion, as well as a negative correlation between the 
early type galaxy content and the group major axis. These correlations
indicate that the cause of the group morphology-velocity dispersion
trend should be attributed mostly to the dynamical evolution of structures and not to
their orientation with respect to the line-of-sight.

Our final conclusion, based on all available evidence regarding the observed
group morphology-dynamics and morphological content-dynamics relations, 
is that the groups of galaxies in the local universe
do not constitute a family of objects in dynamical equilibrium, but rather a family of cosmic
structures that are presently at various stages of their virialization process. We also
expect that the observed group morphology-dynamics correlations are affected by the group
orientation with respect to the line-of-sight, but such an effect works in the same direction
%, as far as the previously mentioned correlations are concerned, 
as the virialization process.

\section* {Acknowledgments}
MP acknowledges funding by CONACyT grant 2005-49878. This 
research has made use of the NASA/IPAC 
Extragalactic Database (NED) which is operated by the Jet Propulsion
 Laboratory, California Institute of Technology, under contract with 
the National Aeronautics and Space Administration.
We thank Cinthia Ragone-Figueroa, Gary Mamon and Manuel Merch\'an 
for useful suggestions and comments.

\newpage

\begin{table}[]
\caption[]{The Pearson correlation coefficents $R$ and 
corresponding random probabilities ${\cal P}$ for the 2MASS-HDC 
and Tago-SDSS group samples with $9\le n_m\le 12$.}
\tabcolsep 8 pt
\begin{tabular}{lccccccc} \\ \hline
  catalog& $\#$  &\multicolumn{2}{c}{$q-\sigma_v$}&
\multicolumn{2}{c}{$b-\sigma_v$}   & \multicolumn{2}{c}{$a-\sigma_v$}\\ \hline
              &   &  R &${\cal P}$&  R   &${\cal P}$&  R   &${\cal P}$ \\ \hline
2MASS-HDC  &  59 & 0.396 & 0.0019 & 0.112 & 0.398  & -0.361 & 0.0046  \\ \hline
Tago-SDSS  & 140 & 0.260 & 0.0019 &  -0.05 & 0.557 & -0.380 & $3\times10^{-6}$ 
\\\hline
\end{tabular}
\end{table}

\begin{table}[]
\caption[]{The Pearson correlation coefficents $R$ and 
corresponding random probabilities ${\cal P}$ for the 2MASS-HDC 
and Tago-SDSS group samples in different mass ranges.}
\tabcolsep 8 pt
\begin{tabular}{lccccccc} \\

     \multicolumn{8}{c}{2MASS-HDC}  \\ \hline
  $\log M/M_{\odot}$ range& $\#$  &\multicolumn{2}{c}{$q-\sigma_v$}&
\multicolumn{2}{c}{$b-\sigma_v$}   & \multicolumn{2}{c}{$a-\sigma_v$}\\ \hline
              &   &  R   &${\cal P}$&  R   &${\cal P}$&  R   &${\cal P}$ \\ \hline
    12-13     & 7 & 0.71 & 0.07  & 0.78 & 0.036   & 0.61 & 0.14     \\
    13-13.5   & 18& 0.49 & 0.04  & 0.31 & 0.21    &-0.34 & 0.16     \\
    13.5-13.75& 23& 0.48 & 0.02  &-0.54 & 0.008   &-0.85 & $<10^{-6}$ \\
    13.75-14  & 29& 0.06 & 0.74  &-0.55 & 0.002   &-0.84 & $<10^{-6}$
    \\ \hline \\ 

     \multicolumn{8}{c}{Tago-SDSS}  \\ \hline
  $\log M/M_{\odot}$ range& $\#$  &\multicolumn{2}{c}{$q-\sigma_v$}&
\multicolumn{2}{c}{$b-\sigma_v$}   & \multicolumn{2}{c}{$a-\sigma_v$}\\ \hline
              &   &  R   &${\cal P}$&  R   &${\cal P}$&  R   &${\cal P}$ \\ \hline
    12-13     & 78& 0.24 & 0.032      & -0.19 & 0.09      & -0.54 & $<10^{-6}$  \\
    13-13.25  & 70& 0.33 & 0.006     & -0.35 & 0.003     & -0.79 & $<10^{-6}$  \\
    13.25-13.5& 71& 0.00 & 0.96      & -0.45 & $10^{-5}$  & -0.66 & $<10^{-6}$   \\
    13.5-13.8& 36& 0.64 &$10^{-5}$   & 0.04 & 0.79        & -0.69 &
    $2\times 10^{-6}$  \\ \hline
\end{tabular}
\end{table} 

\end{document}